\documentclass[11pt,a4paper]{article}

\usepackage{indentfirst}
\usepackage{amsfonts}
\usepackage{amssymb}
\usepackage[reqno,intlimits]{amsmath}
\usepackage[polish,english]{babel}

\newtheorem{theorem}{Theorem}

\newcommand{\ud}{\mathrm{d}}

\title{Global dynamics of cosmological scalar fields -- Part I}
\author{Andrzej J.~Maciejewski,\\
Institute of Astronomy,
University of Zielona G\'ora\\
Podg\'orna 50, 65--246 Zielona G\'ora, Poland,
(e-mail: maciejka@astro.ca.wsp.zgora.pl)\\[2ex]
Maria Przybylska,\\
Institut Fourier, UMR 5582 du CNRS,\\
Universit\'e de Grenoble I,\\
100 rue des Maths,\\
BP 74, 38402 Saint-Martin d'H\`eres Cedex, France\\
and\\
Toru\'n Centre for Astronomy,
Nicholaus Copernicus University \\
Gagarina 11, 87--100 Toru\'n, Poland, (e-mail:
mprzyb@astri.uni.torun.pl)\\[2ex]
Tomasz Stachowiak,\\
Astronomical Observatory, Jagiellonian University\\
Orla 171, 30-244 Krak\'ow, Poland, (e-mail:
toms@oa.uj.edu.pl)\\[2ex]
Marek Szyd{\l}owski,\\
Astronomical Observatory, Jagiellonian University\\
Orla 171, 30-244 Krak\'ow, Poland, (e-mail:
uoszydlo@cyf-kr.edu.pl)}

\begin{document}
\maketitle

\begin{abstract}
We investigate the Liouvillian integrability of Hamiltonian systems
describing a universe filled with a scalar field (possibly complex). The tool
used is the differential Galois group approach, as introduced by
Morales-Ruiz and Ramis. The main result is that the generic systems are
non-integrable, although there still exist some values of parameters for which
integrability remains undecided. We also draw a connection with chaos present in
such cosmological models. The first part of the article deals with minimally
coupled fields, and the second treats the conformal couping.
\end{abstract}

\section{Introduction}

Homogeneous and isotropic cosmological models, although very simple,
explain the recent observational data very well
\cite{Riess:1998cb,Perlmutter:1998np}. Their foundation
is the Friedmann-Robertson-Walker (FRW) universe, described by the metric
\begin{equation}
    \ud s^2 = a^2 \left[ - \ud\eta^2 + \frac{\ud r^2}{1-Kr^2} +
    r^2\ud^2\Omega_2 \right],
\end{equation}
where $a$ is the scale factor, $\ud^2\Omega_2$ is the line element on a
two-sphere, and we chose to use the conformal time $\eta$. As the scale factor
depends only on time, so do all the quantities describing the matter filling
such a universe. These include the density, pressure, a scalar or vector field,
and of course the cosmological constant (with a rather trivial dependence on
$\eta$).

That last element provides an explanation for the current accelerating
expansion of the universe ($\Lambda$ Cold Dark Matter model
\cite{Copeland:2006wr}),
but a better solution still is sought for. A scalar
field which could model a matter with negative pressure is extensively used for
that purpose. A Bose-Einstein condensate of a possibly axionic field is one
idea \cite{Dymnikova:2001ga}, a phantom field violating the energy principle is another
\cite{Caldwell:1999ew}. Finally, scalar field could also be the mechanism behind
inflation \cite{Linde:1981mu}.

A universe filled with only one component seems simple enough but it is not the
case here. Chaos has been studied in conformally coupled fields by means of
Lyapunov exponents, perturbative approach, breaking up of the KAM tori
\cite{Calzetta:1992bv,Bombelli:1997ut},
non-integrability with Painleve property \cite{Helmi:1997mj}. Chaotic
scattering has been found in minimally coupled fields \cite{Cornish:1997ah},
as has dynamics \cite{Monerat:1998tw}.

The Hamiltonian of these systems has indefinite kinetic energy part, and to
cast it into a positive-definite form a transition into imaginary variables is
used. It has been done for a conformally coupled field \cite{Joras:2003dn},
but some authors argue that \cite{Motter:2002jj} there are physical limitations
which forbid extending the solutions through singularities such as $a=0$, and
an imaginary scale factor seems even less realistic.

Despite the fact, that systems of the considered type were called
non-integrable, there was no rigorous proof of that proposition. However, the
Liouvillian integrability can be studied successfully, as we try to show in
these two articles.

\section{Liouvillian integrability}
Here we only describe the  basic notions and facts concerning the 
Morales-Ramis approach 
following \cite{Morales:99::,Ziglin:83::a,Ziglin:83::b} and differential 
Galois theory following \cite{Beukers:92::,Singer:90::}. A more detailed and 
complete presentation of Morales-Ramis theory can be found 
in \cite{Churchill:96::b,Morales:99::,Morales:01::b1} and of differential 
Galois theory in \cite{Kaplansky:76::,Morales:99::,Put:02::}.

Let us  consider a system of differential equations
\begin{equation}
\label{eq:ds}
\dot x = v(x), \qquad t\in\mathbb{C}, \quad x\in M, 
\end{equation}
defined on a complex $n$-dimensional manifold $M$.  If $\varphi(t)$ is
a non-equilibrium solution of \eqref{eq:ds}, then the maximal analytic
continuation of $\varphi(t)$ defines a Riemann surface $\Gamma$ with
$t$ as a local coordinate.  Together with system \eqref{eq:ds} we can
also consider linear variational equations (VEs) restricted to $T_{\Gamma}M$,
i.e.
\begin{equation}
\label{eq:vds}
 \dot \xi = T(v)\xi, \quad T(v)=\dfrac{\partial
v(\varphi(t))}{\partial x},\qquad \xi \in T_\Gamma M.
\end{equation}
We can always reduce the order of this system by one considering 
induced (also linear) system on the normal bundle $N:=T_\Gamma
M/T\Gamma$ of  $\Gamma$  \cite{Kozlov:96::}
\begin{equation}
\label{eq:gnve}
 \dot \eta = \pi_\star(T(v)(\pi^{-1}\xi)), \qquad \eta\in N.
\end{equation}
Here $\pi: T_\Gamma M\rightarrow N$ is the projection.  Obtained in
this way, a system of $s=n-1$ equations is called the normal variational
equations (NVEs). If system \eqref{eq:ds} is Hamiltonian then $n=2m$ and 
the order of VEs can be reduced by two. Namely using a first integral --
 the Hamiltonian we can restrict system  \eqref{eq:ds} to the level $H(x)=E$.
Next the obtained system of $n-1$ differential equations is reduced once more
 by one using the described above reduction on the normal bundle. As 
result we obtain NVEs of order $s=2(m-1)$ which are additionally  Hamiltonian.
 
An analytic
continuation of $\Xi(t)$ -- a matrix of fundamental solutions -- along a loop
$\gamma$ gives rise a new matrix of
fundamental solutions $\widehat\Xi(t)$
which not necessarily coincides with $\Xi(t)$.  However, solutions of
a linear system form $n$ dimensional linear space, so we have
$\widehat\Xi(t) = \Xi(t)M_\gamma$, for certain nonsingular matrix
$M_\gamma\in\mathrm{GL}(n,\mathbb{C})$ which is called the monodromy matrix.
Taking into account loops in different homotopy classes, this leads to the
monodromy group $\mathcal M$.

The monodromy group of NVEs was used by
Ziglin to formulation   
 necessary conditions for the existence of maximal number of first
integrals (without involutivity property) for analytic Hamiltonian
systems \cite{Kozlov:96::,Ziglin:83::a,Ziglin:83::b}. The Ziglin method was 
applied to an analysis a number various system: e.g the rigid body 
\cite{Ziglin:83::b},
the Henon-Heiles system \cite{Ziglin:83::b} etc. However its range of possible
 applications is strongly restricted by difficulties in calculation the
 monodromy groups. These groups are known for a very limited number of 
equations.

Recently Morales-Ruiz and Ramis generalised the Ziglin results
replacing the monodromy group $\mathcal{M}$ by the 
differential Galois group $\mathcal{G}$ of NVEs. There are two important
 advantages of this replacement
\begin{itemize}
\item always $\mathcal{M}\subset\mathcal{G}$ and integrability theorems 
formulated by means of $\mathcal{G}$ are ``stronger'',
\item $\mathcal{G}$ has better algebraic properties and there are tools to
 calculations
 $\mathcal{G}$ for some classes of differential equation.
\end{itemize}

The main problem of differential Galois theory is a problem of solvability of
linear differential equations and systems of linear differential equations
in a class of ``known'' functions.

Let $F$ be a differential field of characteristics zero, i.e. a field with an 
additive operation $':F\to F$ called derivation satisfying the Leibnitz rule. 
In this paper we meet two examples: $F=\mathbb{C}(z)$ -- the set of rational 
functions with coefficients in  
$\mathbb{C}$ and $\mathbb{M}(z)$ - the set of meromorphic
 functions on Riemann surface $\Gamma$. In both examples the derivation is 
the standard 
differentiation $\mathrm{d}/\mathrm{d}z$. The kernel of derivation is a 
sub-field of $F$ called the sub-field of constants and we denote it as $F'$. 
In both our examples $F'=\mathbb{C}$. 

Let $F$ and $L$ are differential fields. The field $L$ is a differential field extension
of $F$ if the derivation on $L$ restricted to $F$ coincides with that of $F$ 
and we denote $F\subset L$.

What we consider is a system of linear equations
\begin{equation}
X'=AX,\qquad '=\dfrac{\mathrm{d}}{\mathrm{d}z},
\label{systemek}
\end{equation}
where $A$ is a matrix of dimension $s$ with entries in $F$ and 
$X=[x_1,\ldots,x_s]^T$. This system can be always transformed to the form
\begin{equation}
y^{(s)}+a_{s-1}y^{(s-1)}+\ldots+a_1y'+a_0y=0i,\qquad a_i\in F.
\label{higher}
\end{equation}
There is also a direct formulation of differential Galois theory for systems 
of linear
differential equations but a majority of literature concerns a linear 
differential equation of degree $s$. For this reason we transformed system 
\eqref{systemek} into \eqref{higher}.

Let $y_{(1)},\ldots,y_{(s)}$ are linearly independent over $F'$ solutions of 
\eqref{higher}. Usually $y_i\not\in F$ and we construct the smallest 
differential extension $K$ containing all solutions of \eqref{higher} and
 such that
$K'=F'$. It is called a Picard-Vessiot extension of $F$. We can imagine $K$ as
 a set $F$ with variables 
$z,y_{(1)},\ldots,y_{(s)},y_{(1)}',\ldots,y_{(s)}',\ldots,y_{(1)}^{(s)},
\ldots,y_{(s)}^{(s)}$. If $F$ has the characteristic zero and $F'$ is 
algebraically closed, then a Picard-Vessiot extension exist and is unique up
 to differential isomorphism.

Next we precise what we understand by ``known'' functions. A solution $y$ of
\eqref{higher} is
\begin{enumerate}
\item \emph{algebraic} over $F$ if $y$ satisfies a polynomial
  equation with coefficients in $F$,
\item \emph{primitive} over $F$ if $y'\in F$, i.e., if
  $y=\int a$, for certain $a\in F$,
\item \emph{exponential}  over $F$ if $y'/y\in F$, i.e., if
  $y=\exp\int a$, for certain $a\in F$.
\end{enumerate}
We say that a differential field $L$ is a Liouvillian extension of $F$
if it can be obtain by successive extensions
\[
F=K_0\subset K_1 \subset \cdots  \subset K_m = L,
\]
such that $K_i=K_{i-1}(y_i)$ with $y_i$ either algebraic,
primitive, or exponential over $K_ {i-1}$. Our vague notion ``known''
functions means Liouvillian function.  We say that~\eqref{higher} is
solvable if its  Picard-Vessiot extension is a Liouvillian
extension.

Using this definition we cannot check solvability of \eqref{higher} in a class
 of Liouvillian functions because
in general we do not know explicit forms of solution. But solvability 
of \eqref{higher}
translates into properties  of the differential Galois group of \eqref{higher}.   
This group can be defined as follows. For
Picard-Vessiot extension $K\supset F$ we consider all automorphisms
of $K$ (i.e. invertible transformations of $K$ preserving field
operations) which commute with differentiation. An automorphism
$g:K\rightarrow K$ commute with differentiation if $g(a')=(g(a))'$,
for all $a\in K$. The set of all such automorphisms we denote $\mathcal{A}$.
Let us note that automorphisms $\mathcal{A}$ form a group. The differential
Galois group (DGG) $\mathcal{G}$ of extension $K\supset F$, is, by definition,
a subgroup of $\mathcal{A}$ such that it contains all automorphisms $g$ which 
do not change elements of $F$, i.e., for $g\in\mathcal{G}$ we have $g(a)=a$
for all $a\in F$.   

Equivalent linear systems of differential 
equations have the same differential Galois groups \cite{Put:02::} and 
differential Galois group of system \eqref{systemek} is isomorphic to a 
differential Galois 
group of associated equation \eqref{higher} of order $s$.

Let us list basic facts about the differential Galois group
\begin{itemize}
\item  $\mathcal{G}$ is an
algebraic subgroup of $\mathrm{GL}(s,F')$. Thus, it is a union of
disjoint connected components. One of them containing the identity is
called the identity component of $\mathcal{G}$ and is denoted by 
$\mathcal{G}^0$.
\item $\mathcal{M}\subset\mathcal{G}$ and in a case when \eqref{higher} is Fuchsian, then $\mathcal{M}$ is dense in Zariski topology in $\mathcal{G}$.
\item Every solution of equation \eqref{higher} is Liouvillian iff
  $\mathcal{G}^0$ conjugates to a subgroup of triangular group.
  This is the Lie-Kolchin theorem.
\end{itemize}

Morales-Ruiz and Ramis showed a connection between integrability of Hamiltonian
systems and properties of the identity component of the differential Galois 
group of NVEs \cite{Morales:99::,Morales:01::b1}.
\begin{theorem}
\label{thm:MR}
  Assume that a Hamiltonian system is meromorphically integrable in
  the Liouville sense in a neighbourhood of a particular solution. 
Then the identity component of the differential Galois
  group of NVEs is Abelian.
\end{theorem}
Application of the Morales-Ramis theory consists of 
\begin{itemize} 
\item finding of particular solution(s), 
\item writing of variational equations and separating  normal variational equations, 
\item checking of the identity component of differential Galois group 
 of normal variational equations. 
\end{itemize} 
In applications, the last step is the most difficult and successful work 
depends on the form of NVEs.

Sometimes the application of variational equations to integrability studies is 
insufficient. Then, often, variational equations of higher orders are useful.
For their definition and applications in integrability theory consult
for example \cite{Maciejewski:2005}.

\section{Minimally coupled field's setup}

The standard, general action for a complex field $\psi$ is given as
\begin{equation}
    \mathcal{I} = \frac{c^4}{16\pi G}\int \left[\mathcal{R} - 2\Lambda -
    \frac12\left(\nabla_{\alpha}\bar{\psi}\nabla^{\alpha}\psi +
    \frac{m^2}{\hbar^2}|\psi|^2\right)\right]\sqrt{-g}\,\ud^4\boldsymbol{\rm x},
\end{equation}
where $\mathcal R$ is the Ricci scalar, $\Lambda$ the cosmological constant,
and $m$ the so called mass of the field. In the FRW model the metric depends
only on time, and so must the field. We choose to use the conformal time
$\eta$, so that the Lagrangian becomes
\begin{equation}
    \mathcal L = 6(a''a + Ka^2) + \frac12 |\psi'|^2a^2 -
    \frac{m^2}{2\hbar^2}a^4|\psi|^2 - 2\Lambda a^4,
\end{equation}
with the prime denoting the derivative with respect to time, and $K$ the index
of curvature. We also dropped a coefficient which includes some physical
constants and the part of the action related to the spatial integration.

Next we subtract a full derivative $6(a'a)'$, and use the polar parametrisation
for the scalar field $\psi = \sqrt{12}\,\phi \exp(i\theta)$ to get
\begin{equation}
    \mathcal L = 6(Ka^2 - a'^2) + 6a^2(\phi'^2+\phi^2\theta'^2) -
    6\frac{m^2}{\hbar^2}a^4\phi^2- 2\Lambda a^4,
\end{equation}
and obtain the Hamiltonian
\begin{equation}
    H = \frac{1}{24}\left(\frac{1}{a^2}p_{\phi}^2 - p_a^2 +
    \frac{1}{a^2\phi^2}p_{\theta}^2\right) - 6Ka^2 + 2\Lambda a^4 +
    6\frac{m^2}{\hbar^2}a^4\phi^2.
\end{equation}
Since $\theta$ is a cyclical variable, the corresponding momentum is
conserved so we substitute $p_{\theta}^2=2\omega^2$. To make all the
quantities dimensionless, we make the following rescalings
\begin{equation}
    m^2\rightarrow m^2\hbar^2|K|,\quad
    \Lambda\rightarrow 3L|K|,\quad \omega^2\rightarrow72\omega^2|K|,
    \quad p_a^2\rightarrow 72p_a^2|K|,\quad p_{\phi}^2\rightarrow
    72p_{\phi}^2|K|.
\end{equation}
There is no need of changing the variables $a$ and $\phi$ along with their
momenta, as this is really changing the time variable $\eta$, and thus the
derivatives to which the momenta are proportional. This result also in
dividing the whole Hamiltonian by $6\sqrt2|K|$ to yield
\begin{equation}
    \sqrt2 H = \frac12 \left(-p_a^2 + \frac{1}{a^2}p_{\phi}^2 \right) -
    \frac{K}{|K|} a^2 + L a^4
    + m^2\phi^2 a^4 +\frac{\omega^2}{a^2\phi^2}.
\end{equation}
If the spatial curvature is zero, any of the other dimensional constants can be
used for this purpose, so
without the loss of generality we take the right-hand side to be the new
Hamiltonian
\begin{equation}
    H = \frac12 \left(-p_a^2 + \frac{1}{a^2}p_{\phi}^2 \right) - k a^2 + L a^4
    + m^2\phi^2 a^4 +\frac{\omega^2}{a^2\phi^2}, \label{MainHam}
\end{equation}
and in all physical cases, $k\in\{-1,0,1\}$, $\omega^2\geqslant 0$,
$m^2\in\mathbb R$, $L\in\mathbb R$, $H=0$. We extend the analysis somewhat
assuming that the Hamiltonian might be equal to some non-zero constant
$E\in\mathbb R$. We will later see, that our analysis
includes also the possibility of these coefficients being complex.

Note that for a massless field, the system is already solvable, as shown in
appendix A.

From this point on, we take $\omega=0$, which means the phase is constant.
Since the model has U(1) symmetry, we can always make such a field real with a
rotation in the complex $\psi$ plane.

Under this assumption the Hamilton's equations of system (\ref{MainHam}) are
\begin{equation}
\begin{aligned}
    \dot{a} &= -p_a, &&&
    \dot{p}_a &= 2k a - 4a^3(L +m^2\phi^2) + \frac{1}{a^3}p_{\phi}^2,\\
    \dot{\phi} &= \frac{1}{a^2}p_{\phi}, &&&
    \dot{p}_{\phi} &= -2m^2 a^4\phi.
\end{aligned} \label{MainEq}
\end{equation}

We note that there is an obvious particular solution, which describes an empty
universe: $\phi=p_{\phi}=0$, $a=q$, $p_a=-\dot{q}$. Thanks to the energy integral
$E=\frac12\dot{q}^2+kq^2-Lq^4$, it can be
identified with an appropriate elliptic function.

%
%
\section{Non-integrability in the $\Lambda=0$ case}
The system now has the following form
\begin{equation}
\begin{aligned}
    \dot{a} &= -p_a, &&&
    \dot{p}_a &= 2k a - 4m^2a^3\phi^2 + \frac{1}{a^3}p_{\phi}^2,\\
    \dot{\phi} &= \frac{1}{a^2}p_{\phi}, &&&
    \dot{p}_{\phi} &= -2m^2 a^4\phi.
\end{aligned}
\end{equation}
Using the aforementioned particular solution, for which the
constant energy condition becomes $E=\frac12\dot{q}^2+kq^2$, we have as the
variational equations
\begin{equation}
\left( \begin{array}{c}
\dot{a}^{(1)}\\
\dot{p}_a^{(1)}\\
\dot{\phi}^{(1)}\\
\dot{p}_{\phi}^{(1)}
\end{array} \right) = \left( \begin{array}{cccc}
0 & -1 & 0 & 0 \\
2k & 0 & 0 & 0 \\
0 & 0 & 0 & \frac{1}{q^2} \\
0 & 0 & -2m^2q^4 & 0
\end{array} \right)
\left( \begin{array}{c}
a^{(1)}\\
p_a^{(1)}\\
\phi^{(1)}\\
p_{\phi}^{(1)} \end{array} \right)
\end{equation}
The normal part of the above system, after eliminating the momentum variation
$p_{\phi}^{(1)}$, and writing $x$ for $\phi^{(1)}$, is
\begin{equation}    
    q \ddot{x} + 2\dot{q}\dot{x} + 2m^2q^3x = 0,
\end{equation}
which we further simplify like before by taking $z=q$ as the new independent
variable, and using the energy condition to get
\begin{equation}
    z(E-kz^2)x''+(2E-3kz^2)x'+m^2z^3 x = 0 \label{RatVarL0}
\end{equation}
We check the physical $E=0$ hypersurface. This requires
$k\ne 0$ for otherwise the special solution would become an equilibrium point.
Introducing a new pair of variables
\begin{equation}
    w(s)=w\left(2\frac{m}{\sqrt{k}}z\right)=z^{3/2}x(z),
\end{equation}
we finally get
\begin{equation}
    \frac{\ud^2w}{\ud s^2}
    = (\frac14 - \frac{\kappa}{z} + \frac{4\mu^2-1}{4s^2})w,
\end{equation}
with $\mu = \pm 1$, and $\kappa=0$. This is the Whittaker equation, and its
solutions are Liouvillian if, and only if,
$(\kappa+\mu-\frac12,\kappa-\mu-\frac12)$ are integers, one of them being
positive and the other negative \cite{Morales:99::}. As this is not the case here,
this finishes the proof for $k\ne0$.

Non-integrability on one energy hypersurface means no global integrability, for
the existence of another integral for all values of $E$ would imply its
existence on $E=0$. However, there might exist additional integrals for only
some, special values of the energy. It is straightforward to check with the use
of Kovacic's algorithm \cite{Kovacic:86::}, that this is not true here. In cases 1
and 2, there is no appropriate integer degree of polynomial needed for the
solution, and case 3 cannot hold, because of the orders of the singular points
of the equation.

If $k=0$, a change of the dependent variable to $w(z)=zx(z)$, reduces 
equation (\ref{RatVarL0}) to
\begin{equation}
    Ew''+m^2z^2w=0,
\end{equation}
which is known not to posses Liouvillian solutions \cite{Kovacic:86::}.

We notice that when $L=E=k=0$, the system can be reduced to a two-dimensional
one. In fact, the reduction is still possible when $L\ne 0$, so we choose to
present in the next section.

%
%

\section{Non-integrability in the $\Lambda\ne 0$ case}

We use the nonzero constant $L$ to rescale the system as follows
\begin{equation}
\begin{aligned}
    a &= \frac{q_1}{\sqrt{L}} ,& p_a &= \frac{p_1}{\sqrt{L}},\\
    \phi & = q_2,& p_{\phi} &= \frac{p_2}{L},
\end{aligned} \label{Scaling}
\end{equation}
so that the equations become\begin{equation}
\begin{aligned}
    \dot{q}_1 &= -p_1, &&&
    \dot{p}_1 &= 2k q_1 - 4q_1^3(1 +b q_2^2) + \frac{1}{q_1^3}p_2^2,\\
    \dot{q}_2 &= \frac{1}{{q_1}^2}p_2, &&&
    \dot{p}_2 &= -2bq_2 q_1^4,
\end{aligned} \label{scaled_eq}
\end{equation}
where $b=m^2/L$. The energy integral, for the previously defined particular
solution, now reads $\mathcal{E} = E L = \frac12 \dot{q}^2 + k q^2 - q^4$,
where $q$ has been rescaled according to (\ref{Scaling}).

As before, we are interested in the variational equations, which read
\begin{equation}
\left( \begin{array}{c}
\dot{q}_1^{(1)}\\
\dot{p}_1^{(1)}\\
\dot{q}_2^{(1)}\\
\dot{p}_2^{(1)}
\end{array} \right) = \left( \begin{array}{cccc}
0 & -1 & 0 & 0 \\
2(k-6q^2) & 0 & 0 & 0 \\
0 & 0 & 0 & \frac{1}{q^2} \\
0 & 0 & -2bq^4 & 0
\end{array} \right)
\left( \begin{array}{c}
q_1^{(1)}\\
p_1^{(1)}\\
q_2^{(1)}\\
p_2^{(1)} \end{array} \right),
\end{equation}
and writing $x$ for $q_2^{(1)}$, and $y$ for $p_2^{(1)}$, the normal part is
\begin{equation}
\begin{aligned}
    \dot{x} &= \frac{1}{q^2} y,\\
    \dot{y} &= -2 b q^4 x.
\end{aligned}
\end{equation}
Or alternatively
\begin{equation}
\ddot{x}+2\frac{\dot{q}}{q}\dot{x}+2bq^2x = 0.
\end{equation}

\subsection{$\mathcal{E}=0$}

We first pick the particular solution of zero energy, as global integrability
implies integrability for this particular value of the Hamiltonian. It is
important to remember, however, that the converse is not true.

The normal variational equation is cast into rational form by changing the
independent variable to $z=q^2/k$ (for $k\ne0$ which implies $k^2=1$), and
using the energy integral
\begin{equation}
    x'' + \frac{5z-4}{2z(z-1)}x' + \frac{b}{4z(z-1)}x=0, \label{Lambda_rat}
\end{equation}
with the respective characteristic exponents
\begin{equation}
\begin{aligned}
    z &= 0, & \rho &= -1,0\\
    z &= 1, & \rho &= 0,\frac12\\
    z &= \infty, & \rho &= \frac14(3-\sqrt{9-4b}),\frac14(3+\sqrt{9-4b})
\end{aligned}
\end{equation}
By Kimura's theorem \cite{Kimura:69::}, the solutions of equation (\ref{Lambda_rat})
are Liouvillian if, and only if $9-4b=(2p-1)^2$, $p\in\mathbb Z$.
As before, this means that for global integrability this condition must be
satisfied.

For $k=0$ the solution of NVE is $x_{1,2}=q^{-2\rho_{\infty 1,2}}$, and the
reduction to two degrees of freedom is possible, as mentioned before.

\subsection{$E\ne0$}

The special solution, is now directly connected to the Weierstrass $\wp$
function, for if we introduce a new dependent variable $v$ with
\begin{equation}
    q^2 = \frac{1}{2} v + \frac{k}{3},
\end{equation}
the energy integral implies that it satisfies the equation
\begin{equation}
    \dot{v}^2 = 4 v^3 - g_2 v - g_3, \label{weiers_eq}
\end{equation}
where
\begin{equation}
g_2=\frac{16}{3}(k^2-3\mathcal E),\;g_3=\frac{32}{27}k(2k^2-9\mathcal E),
\end{equation}
and the discriminant $\Delta=1024 {\mathcal E}^2 (k^2 - 4\mathcal E)$, which we
take as non-zero to consider the generic case. Thus, taking
$w = q_2^{(1)} q$, and eliminating $p_2^{(1)}$ as before, the normal variational equation
reads
\begin{equation}
    \ddot{w} = [A\wp(t;g_2,g_3) + B]w,
\end{equation}
with $A=2-b$ and $B=-\frac{2}{3}k(1 + b)$. This is the Lam\'e differential
equation, whose Liouvillian solutions are known to fall into three mutually
exclusive cases, which are exactly those of Kovacic's algorithm:
\begin{enumerate}
\item The Lam\'e-Hermite case, with $A=n(n+1)=2-b$, $n\in\mathbb Z_+$. This implies
that $9-4b=(2n+1)^2$.
Since $b$ does not limit the values of $n$ in any other way, we have to 
use the higher variational equations, to prove that the Galois group is
non-abelian. To do that,
it is convenient to change the variables of equations (\ref{scaled_eq}) in the
following way
\begin{equation}
\begin{aligned}
    q_1 &= w_1, & p_1 &= -w_2,\\
    q_2 & = \frac{w_3}{w1}, & p_2 &= w_1 w_4 - w_2 w_3,
\end{aligned}
\end{equation}
so that the variational equations are all Lam\'e's equations
\begin{equation}
\left( \begin{array}{c}
\dot{w}_1^{(1)}\\
\dot{w}_2^{(1)}\\
\dot{w}_3^{(1)}\\
\dot{w}_4^{(1)}
\end{array} \right) = \left( \begin{array}{cccc}
0 & 1 & 0 & 0 \\
A_1\wp(\eta) + B_1 & 0 & 0 & 0 \\
0 & 0 & 0 & 1 \\
0 & 0 & A_2\wp(\eta) + B_2 & 0
\end{array} \right)
\left( \begin{array}{c}
w_1^{(1)}\\
w_2^{(1)}\\
w_3^{(1)}\\
w_4^{(1)} \end{array} \right),
\end{equation}
where $\wp(\eta)$ is the one given by equation (\ref{weiers_eq}), and
\begin{equation}
\begin{aligned}
    A_1 &= 6, & B_1 &= 2k\\
    A_2 &= n(n+1), & B_2 &= \frac23 k(n^2+n-3).
\end{aligned}
\end{equation}

Using the same method as in \cite{Maciejewski:2005}, we find that the fourth order variational
equation's solution involves a logarithm, provided that
$k\ne0$, thus rendering the Lam\'e-Hermite case non-integrable. Note, that HVE
can only be used in the first case of Lam\'e's equation \cite{Morales:99::}.

\item The Brioschi-Halphen-Crawford case, where necessarily $n$ is half an
integer, i.e. $n+\frac12=l\in\mathbb N$, and as before $9-4b=(2n+1)^2=(2l)^2$.

\item The Baldassarri case, with $n+\frac12 \in \frac13 \mathbb Z \cup
\frac14\mathbb Z\cup\frac15\mathbb Z\setminus\mathbb Z$, and
additional algebraic restrictions on $B,g_2$, and $g_3$.
However, inspecting the third case using Kovacic's algorithm, we see that
$9-4b=(2n+1)^2$ must be a square of an integer which is impossible if $n$
belongs to the family mentioned.
\end{enumerate}

\subsection{$E=k=0$}
As mentioned in the $\Lambda=0$ section, we can transform the
system to a two-dimensional one.
In order to do that, time needs to be changed from the conformal to the
cosmological one $\ud\eta \rightarrow \ud t = a\ud\eta$, in the original
equations (\ref{MainEq}). We then take as the
new momenta the Hubble's function and the derivative of $\phi$
\begin{equation}
\begin{aligned}
    h &:= \frac{1}{a}\frac{\ud a}{\ud t} = -\frac{p_a}{a^2}\\
    \omega &:= \frac{\ud\phi}{\ud t} = \frac{p_{\phi}}{a^3}.
\end{aligned}
\end{equation}
(This $\omega$ is not to be confused with the one introduced in section 3.)
Accordingly we have
\begin{equation}
\begin{aligned}
    \frac{\ud a}{\ud t} &= ah, \\
    \frac{\ud\phi}{\ud t} &= \omega, \\
    \frac{\ud h}{\ud t} &= 4L + 4m^2\phi^2-\omega^2-2h^2, \\
    \frac{\ud\omega}{\ud t} &= -2m^2\phi-3\omega h.
\end{aligned}
\end{equation}
Thus, we are left with a dynamical system in the $(h,\phi,\omega)$ space, as $a$
decouples. Furthermore, the energy integral is now
\begin{equation}
    0 = \frac12 a^4(2L + \omega^2 + 2m^2\phi^2 - h^2),
\end{equation}
so for $a(t)$ which is not trivially zero, it gives a first integral on the
reduced space. Choosing an appropriate variable $\alpha$, suggested by the
form of this integral
\begin{equation}
\begin{aligned}
    \phi &= \frac{\sqrt{h^2-2L}}{\sqrt{2}m}\sin(\alpha),\\
    \omega &= \sqrt{h^2-2L}\cos(\alpha),
\end{aligned}
\end{equation}
we finally obtain
\begin{equation}
\begin{aligned}
    \frac{\ud\alpha}{\ud t} &= \sqrt{2}\,m+3h\sin(\alpha)\cos(\alpha),\\
    \frac{\ud h}{\ud t} &= -3(h^2-2L)\cos(\alpha)^2.
\end{aligned}
\end{equation}

The problem of such reduction was
also discussed in \cite{Faraoni:2006sr}. It is argued that there can be no
chaos in this system, but its integrability -- which would be one more first
integral -- remains unresolved.

\section{Conclusions}

The main result of our paper can be summarised as follows.

When $\Lambda=0$, and if the system is meromorphically integrable, then
necessarily  $E=k=0$, in which case it is possible to reduce it to
two-dimensions, and there is no chaos.

When $\Lambda\ne 0$, we consider the physical $E=0$ hypersurface first.
If the system is integrable, then either $k=0$ (reducibility as above), or
$9-4m^2/L=(2n+1)^2$, $n\in\mathbb Z$.

When $\Lambda\ne 0$, and we consider a generic energy hypersurface,
integrability implies that either $9-4m^2/L=(2n+1)^2$ and
$k=0$, or $9-4m^2/L=(2n)^2$ (regardless of the value of $k$), for some
$n\in\mathbb Z$.

These are, however, only necessary and not sufficient conditions, so that the
system might still prove not to be integrable at all. In particular, the
numerical search for chaos suggests both the lack of global first integrals,
and crucial differences in the behaviour of the system for real and imaginary
values of the variables. This might be a clue, that the system might have first
integrals which are not analytic, and thus not prolongable to the complex
domain. A system with similar property was studies by the authors in
\cite{Maciejewski:2005yi}.

It might also prove useful, to investigate the existence of asymptotic first
integrals, but this is more in connection with scattering dynamics of
conformally coupled fields, which we study in the second part of the paper.

\section*{Acknowledgements}
For the second author this research has been partially supported by the
European Community project GIFT  (NEST-Adventure Project no. 5006) and by
Projet de l'Agence National de la Recherche ``Int\'egrabilit\'e r\'eelle et
complexe en m\'ecanique hamiltonienne'' N$^\circ$~JC05$_-$41465. And for the
fourth author, by Marie Curie Host Fellowship MTKD-CT-2004-517186 (COCOS).

\section*{Appendix A. A solvable case}

When $m=0$, the Hamilton-Jacobi equation for the main Hamiltonian
(\ref{MainHam}) will become separable, because it can be written as
\begin{equation}
    Ea^2 = \frac12\left(\frac{\partial W}{\partial\phi}\right)^2
    - \frac12 a^2 \left(\frac{\partial W}{\partial a}\right)^2
    - ka^4 +La^6 +\frac{\omega^2}{\phi^2}, \label{EqHJ}
\end{equation}
with the full generating function $S=W-E\eta$. Assuming $W=A(a)+F(\phi)$,
equation (\ref{EqHJ}) can be solved with
\begin{equation}
\begin{aligned}
    F(\phi) &= \int\sqrt{2\left(J-\frac{\omega^2}{\phi^2}\right)}\ud\phi\\
    A(a) &= \int\sqrt{2\left(La^4-ka^2-E+\frac{J}{a^2}\right)}\ud a.
\end{aligned}
\end{equation}
The first equation of motion can then be deduced from
\begin{equation}
    \frac{\partial W}{\partial E} - \eta =
    \int\frac{\ud a}{\sqrt{2(La^4-ka^2-E+\frac{J}{a^2})}} =
    {\rm const},
\end{equation}
which can be rewritten as
\begin{equation}
    \left(\frac{\ud a}{\ud\eta}\right)^2 =
    2(La^4-ka^2-E+\frac{J}{a^2}).
\end{equation}
Or, introducing a new variable $v=a^2$, as
\begin{equation}
    \left(\frac{\ud v}{\ud\eta}\right)^2 =
    8(Lv^3-kv^2-Ev+J), \label{EqWei}
\end{equation}
so that the general solution is
\begin{equation}
    a^2 = v = \frac{1}{2L}\wp(\eta-\eta_0;g_2,g_3)+\frac{k}{3L},
\end{equation}
where
\begin{equation}
\begin{aligned}
    g_2 &= \frac{16}{3}k^2+16LE \\
    g_3 &= \frac{32}{3}LkE+\frac{64}{27}k^3-32L^2J,
\end{aligned}
\end{equation}
and $\eta_0$ is the constant of integration.
Of course, for $L=0$, equation (\ref{EqWei}) admits solutions in terms of
circular functions.

The equation for $\phi(\eta)$ is the following
\begin{equation}
    \frac{\partial W}{\partial J} =
    \int\frac{\ud\phi}{\sqrt{2(J-\frac{\omega^2}{\phi^2})}} +
    \int\frac{\ud a}{a^2\sqrt{2(La^4-ka^2-E+\frac{J}{a^2})}} =
    {\rm const}, \label{Int}
\end{equation}
which we simplify by using the just obtained solution for $v(\eta)$ to get
\begin{equation}
    {\rm const} = \frac{\sqrt{J\phi^2-\omega^2}}{\sqrt2J} + 
    \int\frac{\ud\eta}{2v}.
\end{equation}
As $v$ is an elliptic function of order two, the second integral can be
evaluated by means of the Weierstrass zeta function to yield
\begin{equation}
    {\rm const} = \frac{\sqrt{J\phi^2-\omega^2}}{\sqrt2J} +
    \frac{1}{4\sqrt{2J}}\big[\zeta(\eta_1)-\zeta(\eta_2)\big]\eta +
    \frac{1}{4\sqrt{2J}}
    \ln\left[\frac{\sigma(\eta-\eta_1)}{\sigma(\eta-\eta_2)}\right],
\end{equation}
where $\eta_{1,2}$, are the zeroes of $v(\eta)$, given by
\begin{equation}
    3\wp(\eta_{1,2};g_2,g_3)=-2k,
\end{equation}
and the constant of integration can be determined from the boundary conditions
on the field $\phi$. Again, for $J=0$, the integrals in (\ref{Int}) reduce to
simpler functions.

\end{document}